\documentstyle[epsfig,rotating]{jeru}
\makeatletter
\let\chapter\hid@chapter
\makeatother
\newcommand{\Pom}{I\!P}                
\newcommand{\Reg}{I\!R}                

\begin{document}
\begin{flushright}
{\bf ETHZ-IPP 97-08}
\end{flushright}

\vspace*{2cm}

\begin{center}
{\LARGE {\bf Soft Interactions and Diffractive Phenomena \footnote{Plenary talk PL2 at the 
Europhysics Conference on High Energy Physics, Jerusalem, August 1997}}}

\vspace*{1cm}

R. A. Eichler \\
Institute for Particle Physics, ETH-Z\"urich, CH 8093 Zurich\\

\end{center}

\authorrunning{R.\,Eichler}
\titlerunning{{\talknumber}: Diffraction and Fragmentation}
 

\def\talknumber{PL2} 

\title{{\talknumber}: Soft Interactions and Diffractive Phenomena }
\author{Ralph\,Eichler
(eichler@particle.phys.ethz.ch)}
\institute{Institute for Particle Physics, ETH-Z\"urich, CH 8093 Zurich}

\maketitle

\begin{abstract}
Recent results on hard diffraction at HERA and the Tevatron are presented.
Charged particle multiplicities in diffraction and differences in multiplicity
in quark and gluon jets measured at LEP are discussed. Spin effects in the 
fragmentation of leading quarks show some interesting features.
\end{abstract}

\section{Total and Elastic Cross Section}
The energy dependence on the center of mass energy $\sqrt{s}$ of all
hadron-hadron total cross sections  is described by  
$$\sigma_{tot}(s)=A_{\Pom}s^{\alpha_{\Pom}(0)-1}+
A_{\Reg}s^{\alpha_{\Reg}(0)-1} \enspace .\label{Xsect}$$ 
with universal exponents
$\alpha_{\Pom}(0)-1$ and $\alpha_{\Reg}(0)-1$ and process dependent
constants $A_{\Pom}$ and $A_{\Reg}$ \cite{Donnachie}.
Regge theory, which  relates poles in a $t$-channel scattering amplitude to 
energy behaviour in the s-channel gives a link to Regge trajectories
$\alpha(t)$. The trajectories have been determined in hadron scattering
and describe also surprisingly well $\gamma \gamma$ and $\gamma p$
scattering. The first term in the total cross section formula dominates the
high energy behaviour and the corresponding Pomeron trajectory
$\alpha_{\Pom}(t)$ is unique in Regge theory with largest intercept and
vacuum quantum numbers: $\alpha_{\Pom}(t)=1.08+0.25~ GeV^{-2}~t.$
Next leading reggeons have approximately degenerate trajectories and carry
the quantum numbers of $\rho,\omega,a_2,f_2$ mesons. They are combined in an
effective trajectory  $\alpha_{\Reg}(t)=0.55+0.9 ~GeV^{-2}~ t.$

\begin{figure}[htbp]

\vspace*{-0.5cm}

\hspace*{1cm}{\bf \Large a)~~~~~~~~~~~~~~~~~~~~~~~~~~ L3} \hspace*{2.5cm} {\bf \Large b)~~~~~~~~~~~~~HERA}

\vspace*{-0.6cm}

\mbox{\epsfig{file=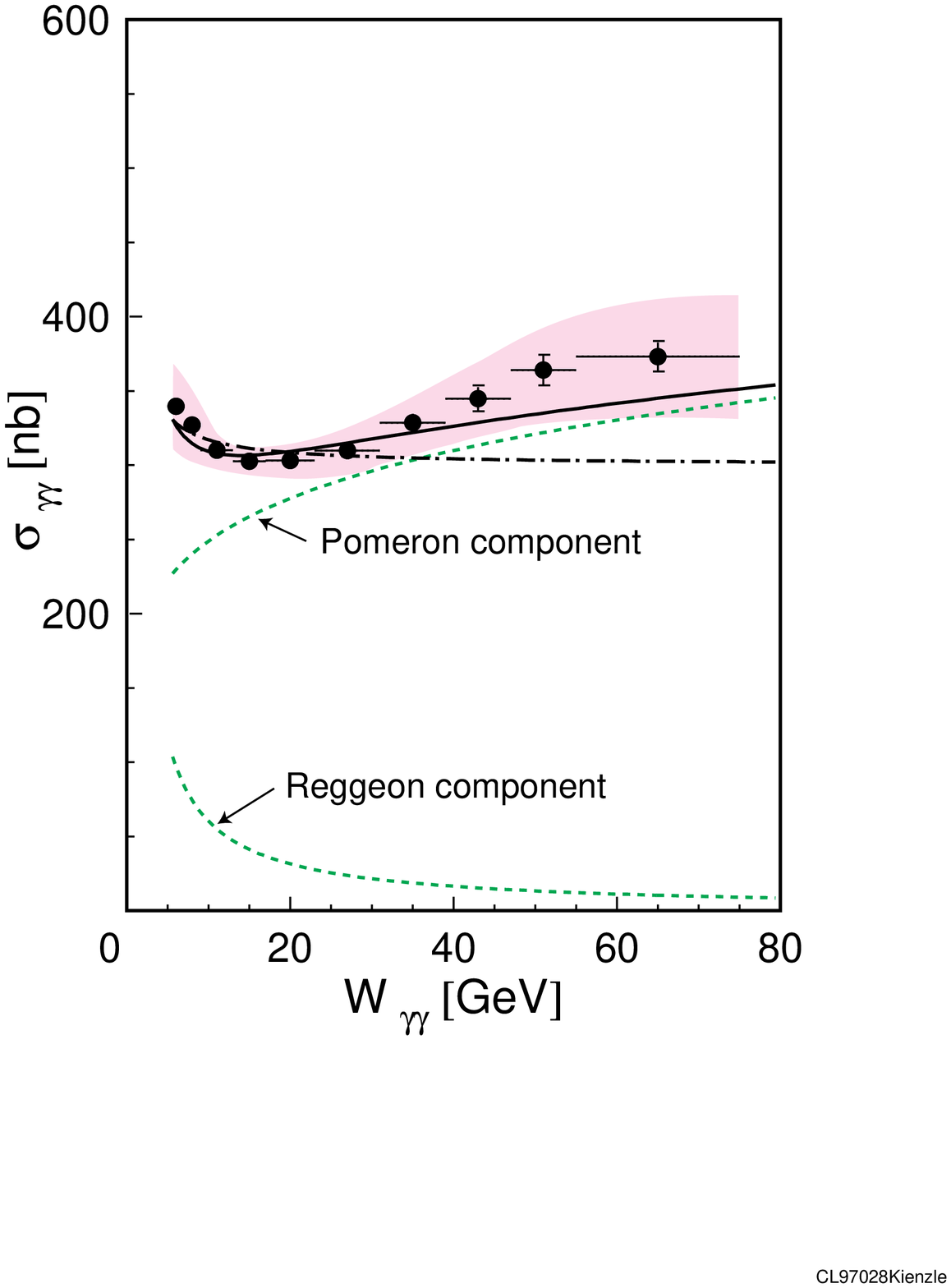,width=0.5\linewidth,clip=}}

\vspace{-7cm}
\hspace*{6.cm}
\mbox{\epsfig{file=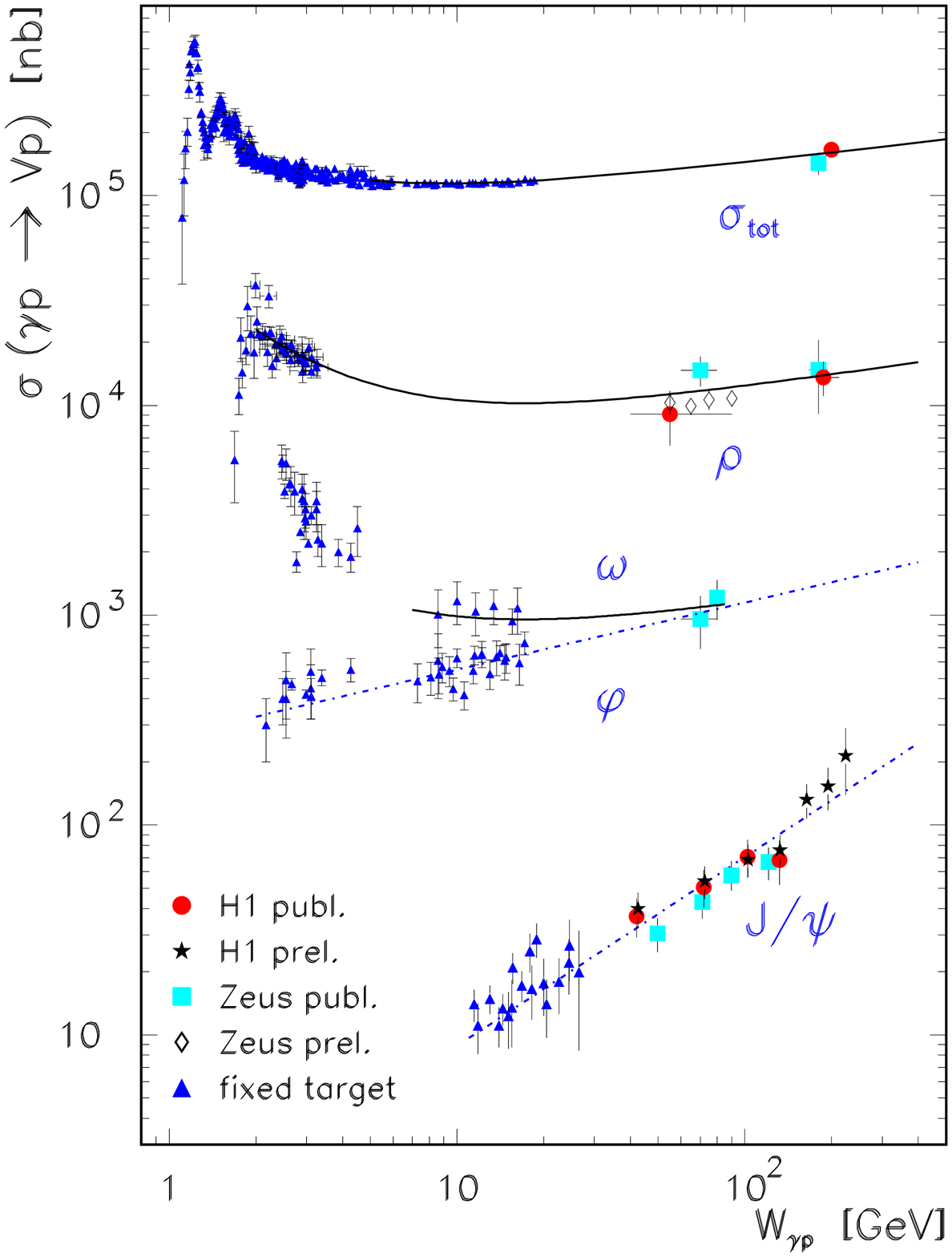,width=0.4\linewidth,clip=}}

\caption[~]{\label{L3} \small a) total $\gamma \gamma$ cross section
versus $\gamma \gamma$ energy $W_{\gamma \gamma}$. The shaded band is the systematic
error, solid line a fit of the sum of Pomeron and Reggeon component.
b) quasireal total and exclusive vector meson $\gamma p$ cross section 
with photon virtuality $Q^2
\sim 0$ vs $\gamma p$ center of mass energy $W_{\gamma p}$. The 
lines are a fit to the data.}    
\end{figure}

Figure \ref{L3}a shows a measurement by the L3 collaboration \cite{L3-coll}
 of the total
$\gamma \gamma$ cross section where the universal exponents have been
assumed and coefficients $A_{\Pom}=173 \pm 7~nb/GeV^{-2}$ and
$A_{\Reg}=519 \pm 125~nb/GeV^{-2}$ were fitted. Figure \ref{L3}b presents
the $\gamma p$ cross section as a function of the $\gamma p$ energy 
$W_{\gamma p}$ at
photon virtualities $Q^2\sim0$. The elastic vector meson production
$\gamma p \rightarrow Vp$ with $V=\rho,\omega,\Phi,J/\Psi$ shows also a
power law behaviour. The elastic cross section is related to the total 
cross section
via the optical theorem $$\frac{d\sigma^{elastic}}{dt}|_{t=0}(\gamma p
\rightarrow Vp) \propto \sigma_{tot}^2 \propto
(W^2)^{2(\alpha_{\Pom}(0)-1)}.$$
 Figure \ref{L3}b suggests that the growth $(W^2)^{2\lambda}$ is steeper
for the heavier vector mesons, namely $\lambda=0.22$ for $J/\Psi$-production
\cite{H1-Psi} compared to the value for $\sigma_{tot}$, namely 
$\lambda=\alpha_{\Pom}(0)-1=0.08$. 
For elastic $\rho$-production the exponent $\lambda$ grows from 0.08 at 
low $Q^2$ to 0.19$\pm$0.07 at $Q^2= 20~GeV^2$ \cite{ZEUS-rho}. Apparently 
$\lambda$ is a function of the 
hardness of the scale (quark mass, $Q^2$) involved in the scattering.

\vspace*{-1.1cm}

\section{Inelastic Diffraction}
\begin{figure}[htbp]

{\bf a) \hspace*{5.cm} b)}

\vspace*{-1.5cm}

\mbox{\epsfig{file=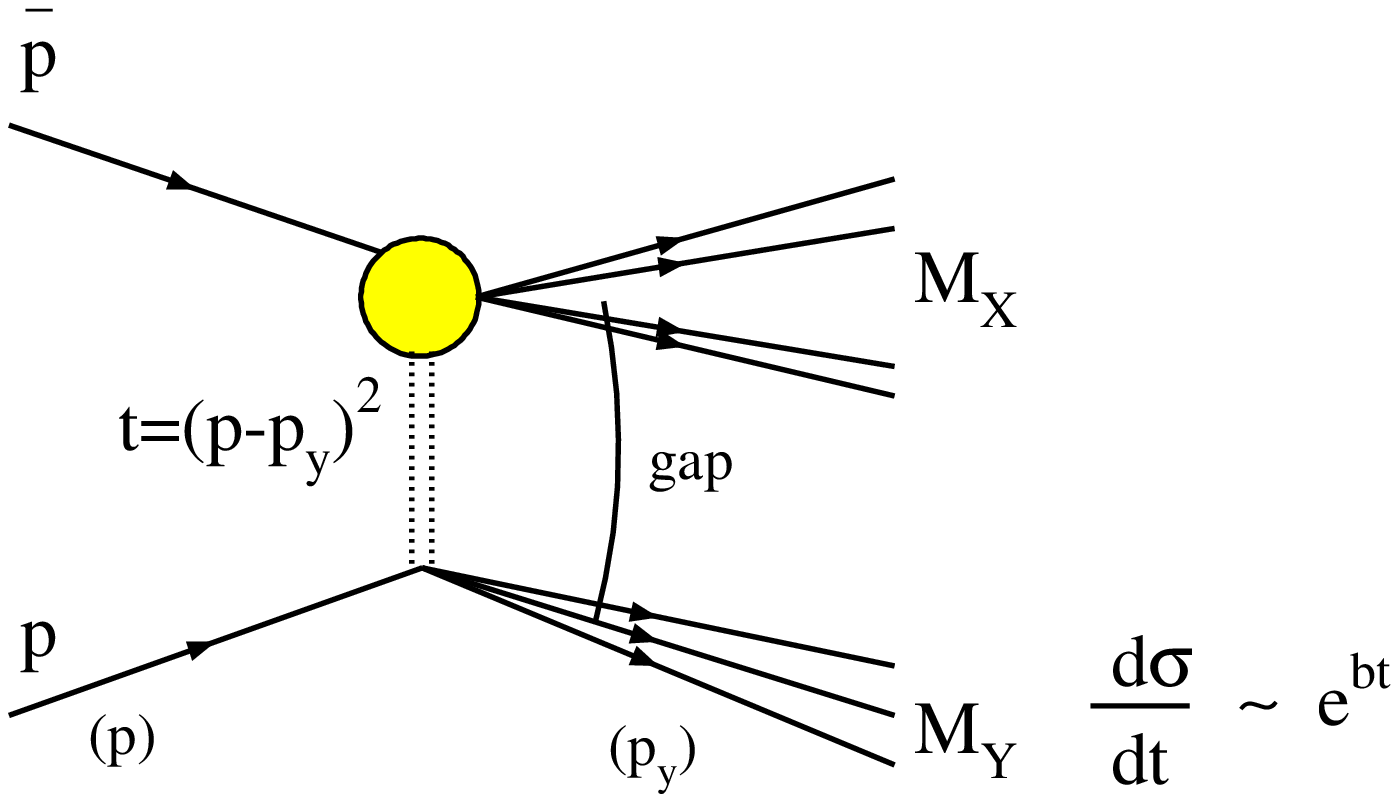,width=0.50\linewidth}}

\vspace*{-4cm}

\hspace*{6cm}
\mbox{\epsfig{file=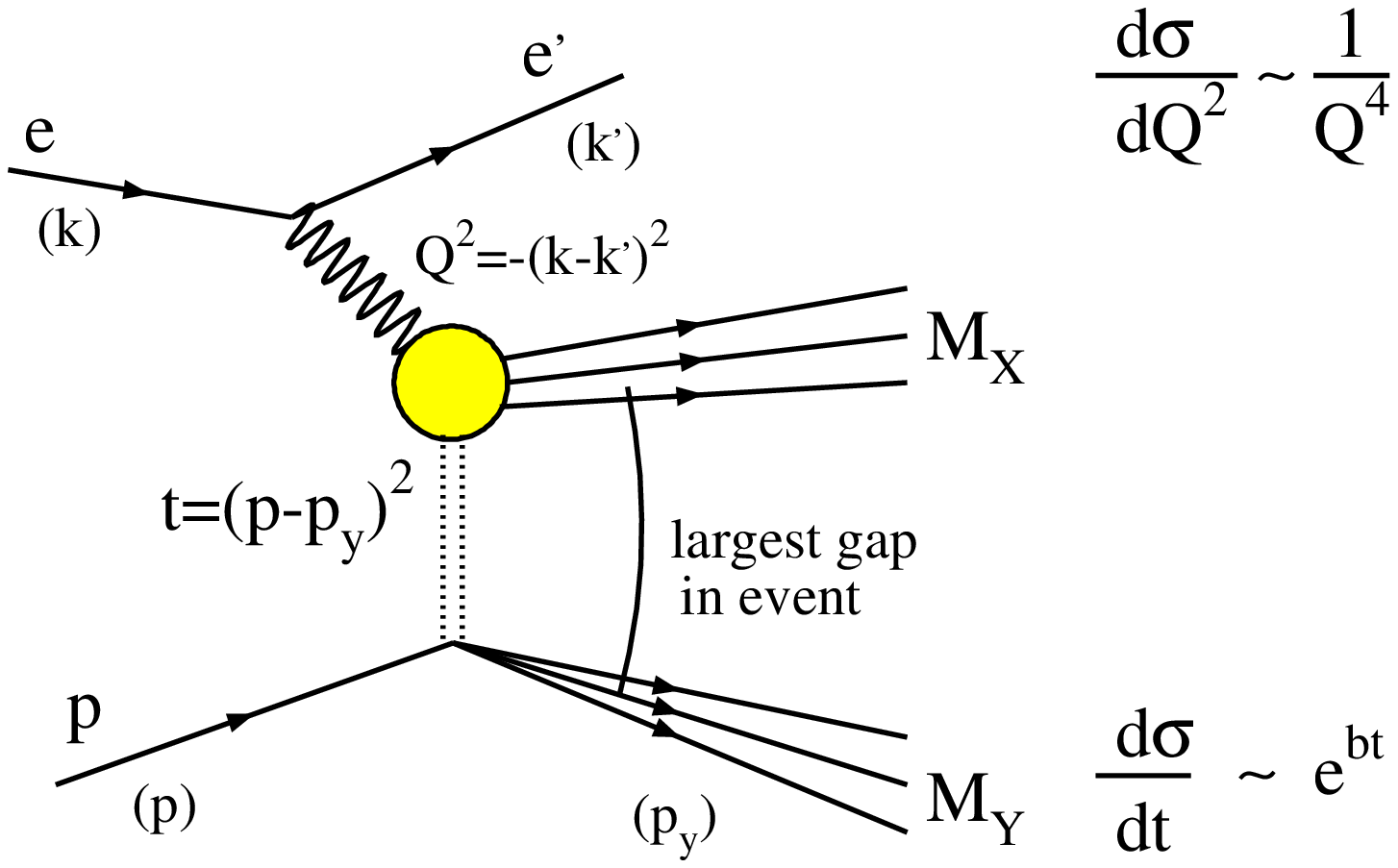,width=0.5\linewidth}}

\caption[~]{\label{diff-picture} \small a)  Diffractive scattering
in $p\bar{p}$ interactions. Hard diffraction is observed in
reactions, where the system $X$ consists of jets.
b) diffractive scattering in deep inelastic $ep$ interactions. Without
detection of the leading proton experimental cuts  constrain $t<1~GeV^2$ and 
the mass $M_Y<1.6~GeV$ (H1), $<4~GeV$ (ZEUS).}      
\end{figure}

Diffractive scattering has been observed in $h-h$ interactions a long
time ago. A typical signature is a forward peaked beam particle which
remains intact or is excited to a small mass $M_Y$ and a rapidity gap 
between it and 
the rest of the final state $X$ (Figure \ref{diff-picture}a).

\begin{figure}[htbp]
\vspace*{-1.cm}
\begin{center}
\mbox{\epsfig{file=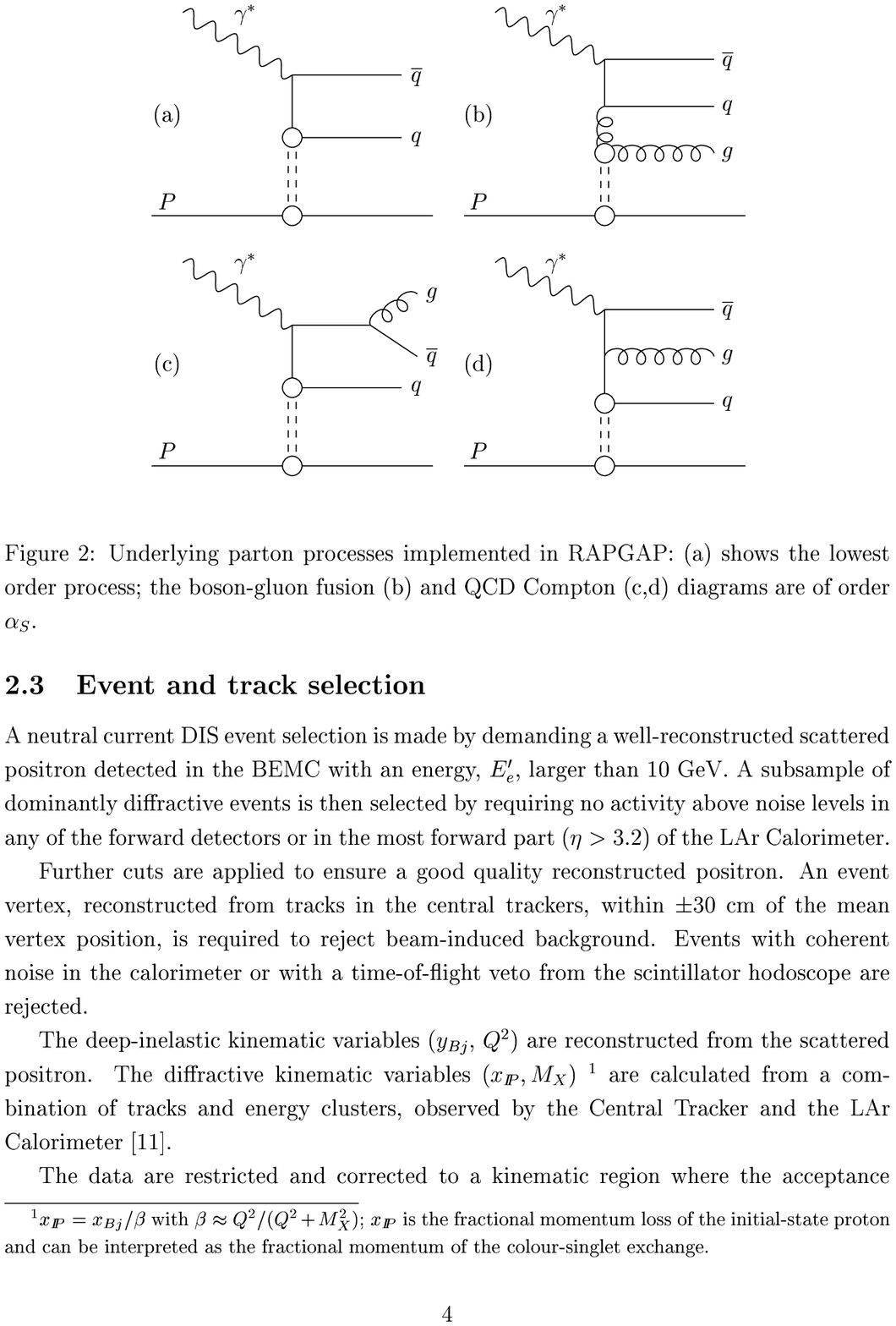,width=0.45\linewidth,clip=}}
\end{center}
\vspace*{-0.5cm}
\caption[~]{\label{Feynman} \small Lowest order Feynman diagrams of
inelastic diffraction in DIS. The dotted double line indicates the
diffractively exchanged object in the infinite momentum frame of the
proton.}
\end{figure}

Diffractive scattering has also been observed in deep inelastic scattering 
(DIS) at HERA with a fraction of
roughly  10\% of total DIS. The kinematics is defined
through the measurement of the scattered
positron, the mass $M_X$ of the system $X$ via the hadronic final
state and the observation of a gap between $X$ and $Y$, 
fig. \ref{diff-picture}b. 
The standard DIS variables are
$q=k-k',~Q^2=-q^2,~W^2=(p+q)^2,~x_{Bj}=q^2/2P\cdot q$
and the additionally measurable variables
$$x_{\Pom} \equiv \xi=\frac{M_X^2+Q^2-t}{W^2+Q^2-m_p^2},
~\beta=\frac{Q^2}{2q\cdot (p-p_Y)}=\frac{x_{Bj}}{\xi}=\frac{Q^2}{M_X^2+Q^2}.$$
The variable $t=(p-p_Y)^2$ is known only if the scattered proton is
detected in the leading proton spectrometer, otherwise data from the forward
detectors only limit $t$ and $M_Y$ (see caption of figure 
\ref{diff-picture}). The
intuitive meaning of $\xi=x_{\Pom}$ and $\beta$ can be best understood in
the infinite momentum frame of the proton. Consider the Feynman diagrams
of Figure \ref{Feynman}. The virtual photon $\gamma^*$ probes partons
in the diffractively exchanged object  which carries a momentum fraction
$x_{\Pom}$ of the proton. These partons carry a fraction $\beta$ of this
object \cite{Schlein}. In the following  it will be shown, that this diffractively
exchanged object can be viewed as in the formula of $\sigma_{tot}$ as a sum
of a Pomeron and a Reggeon term.

\vspace*{-0.5cm}

An equivalent picture of diffraction is seen in the proton rest frame \cite{Bjorken,
Buchmueller,Brodsky,Nikolaev,Soper}.
The Fock states of the photon $|\gamma>=|q\bar{q}>,|q\bar{q}g>,...$ etc are
formed long before the target (Ioffe length $\ell \sim 1/x \sim $several
100 fm) and the quark pairs interact softly through multiple gluon exchange 
with the target. Despite hard scattering at large $Q^2$ the target is
only hit gently. The rapidity gap formation is then a long time scale
process after the quark pair has traversed the proton and the diffractively produced final states
$X$ are expected to be sensitive to the topology and colour structure of the partonic 
fluctuations of the virtual photon.

\subsection{Results from Leading Proton Spectrometer at HERA}
\begin{figure}[htbp]
\begin{center}
\vspace*{-0.7cm}
\mbox{\epsfig{file=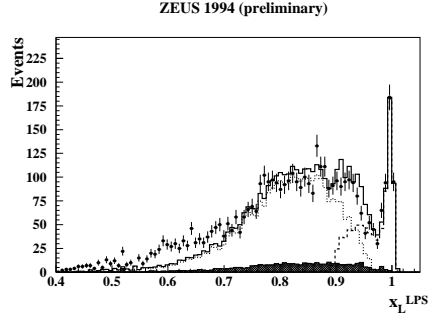,width=0.5\linewidth,clip=}}
\end{center}
\vspace*{-0.5cm}
\caption[~]{\label{LPS} \small Preliminary results from the ZEUS
leading proton spectrometer.}      
\end{figure}

Figure \ref{LPS} shows preliminary results from the ZEUS leading proton
spectrometer \cite{ZEUS-LPS}. The device measures the energy $E_p'$ of the scattered proton
and its transverse momentum $p_t$. The kinematical variables are
defined $ x_L=1-\xi=E_p'/E_p$  and $t=-p_t^2/x_L-m_p^2(1-x_L)^2/x_L.$
A clear diffractive peak at $x_L=1-\xi>0.97$ is observed and the $b$-parameter
($d\sigma/dt\propto e^{bt}$) was fitted and found to be between 4 and 10 depending
on $x_L$ \cite{ZEUS-LPS}. 

\vspace*{-0.9cm}

\subsection{Diffractive Structure Function}
The fivefold differential cross section defines the diffractive structure
function $F_2^{D(5)}$ with five variables \cite{Soper-SF}
$$\frac{d^5\sigma_{ep \rightarrow eXY}}{d\beta dQ^2d\xi dtdM_Y}=
\frac{4\pi \alpha^2}{\beta Q^4}\left(1+(1-y)^2
\right) F_2^{D(5)}(\beta,\xi,Q^2,t,M_Y).$$
To gain statistics the final state proton is not detected and the
variables $M_Y$ and $t$ are not measured. Therefore one defines a new
structure function by integrating over the variables $t$ and $M_Y$.
(H1: $|t|<$ 1 GeV$^2$ and $ M_Y<$ 1.6 GeV,
ZEUS: $|t|<$ 1 GeV$^2$ and $ M_Y<$ 4.0 GeV)
$$\frac{d^3\sigma_{ep \rightarrow eXY}}{d\beta dQ^2d\xi}=
\frac{4\pi \alpha^2}{\beta Q^4}\left(1+(1-y)^2
\right) F_2^{D(3)}(\beta,\xi,Q^2).$$
H1 made an Ansatz for $F_2^{D(3)}$ as a sum of Pomeron exchange and
Reggeon (Meson) exchange \cite{H1-Diff}. Each term factorises into a Pomeron/Meson flux
dependent function of $\xi,t$ and a structure function dependent on
$\beta,Q^2$:
$$F_2^{D(3)}(\beta,\xi,Q^2)=\int dt \left[ \frac{e^{B_{\Pom}t}}
{\xi^{\alpha_{\Pom}(t)-1}}
F_2^{\Pom}(\beta,Q^2,t)+C_M\frac{e^{B_{\Reg}t}}
{\xi^{2\alpha_{\Reg}(t)-1}}
F_2^M(\beta,Q^2,t) \right]$$
plus a possible interference term 
(since the $f_2$-meson and the $\Pom$ have the same C-parity).  
 In the above expression the integral is
 over $|t|<$1 GeV$^2$. 
\begin{figure}[htbp] 
\begin{center}
\vspace*{-0.6cm}
\mbox{\epsfig{file=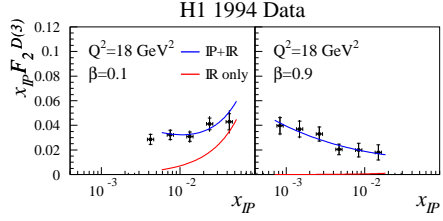,width=0.56\linewidth,clip=}}
\end{center}
\vspace*{-0.6cm}
\caption[~]{\label{fit-F2} \small $x_{\Pom} F_2^{D(3)}(\beta,Q^2)$ as a
function of $x_{\Pom}$ for two bins in $Q^2$ and $\beta$.}       
\end{figure}

\begin{figure}[htbp] 
\begin{center}
\vspace*{-0.99cm}
\mbox{\epsfig{file=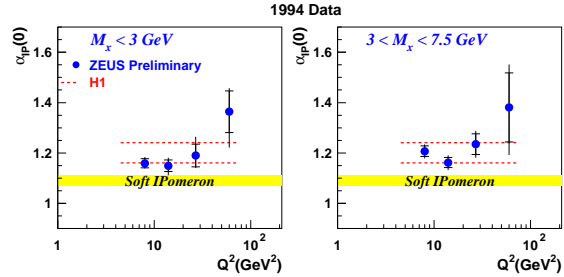,width=0.7\linewidth,clip=}}
\end{center}
\vspace*{-0.5cm}
\caption[~]{\label{alphapom} \small $\alpha_{\Pom}(0)$ vs $Q^2$
in two bins of $M_X$. The band limited by the dotted lines
indicate the range in $\alpha_{\Pom}(0)$ from the H1 analysis.
The shaded band is the value found in soft processes like $\sigma_{tot}$.}        
\end{figure}

The measurements of H1 were fitted in 47 bins in 
$4.5~GeV^2 \le Q^2 \le 75~GeV^2,~0.04\le \beta \le 0.9$
with free parameters $\alpha_{\Pom}(0),\alpha_{\Reg}(0),C_M$.
A very good fit confirms the Ansatz and two such examples are shown in Figure
\ref{fit-F2}. Contributions of the meson term is noticable at 
$\xi=x_{\Pom}>0.01$
and small $\beta$ (large $M_X$). As a result H1 gets \cite{H1-Diff} 
$$\alpha_{\Pom}(0)=1.203\pm
0.02(stat)\pm0.013(syst)^{+0.03}_{-0.035}(model)$$
$$\alpha_{\Reg}(0)=0.50\pm0.11(stat)\pm0.11(syst)\pm0.10(model).$$ 
An analysis of the ZEUS collaboration \cite{ZEUS-Diff} with a fit at
fixed $Q^2$
$$\frac{d\sigma_{ep \rightarrow eXY}^D}{dM_X} \sim
(W^2)^{2\overline{\alpha_{\Pom}}-2}$$  where the non-diffractive
background  has been parametrised and 
subtracted first yields a similar result for 
$\alpha_{\Pom}(0)$ (see Figure \ref{alphapom}).

\vspace*{-0.2cm}

The results of the two experiments are consistent with each other and
the intercept in hard diffraction is larger than $\alpha_{\Pom}(0)=1.08$
determined from $\sigma_{tot}.$

\begin{figure}[htbp] 
{\bf \Large a)} \hspace{5cm} {\bf \Large b)}

\vspace{-1.2 cm}

\hspace{2.5cm}

\mbox{\epsfig{file=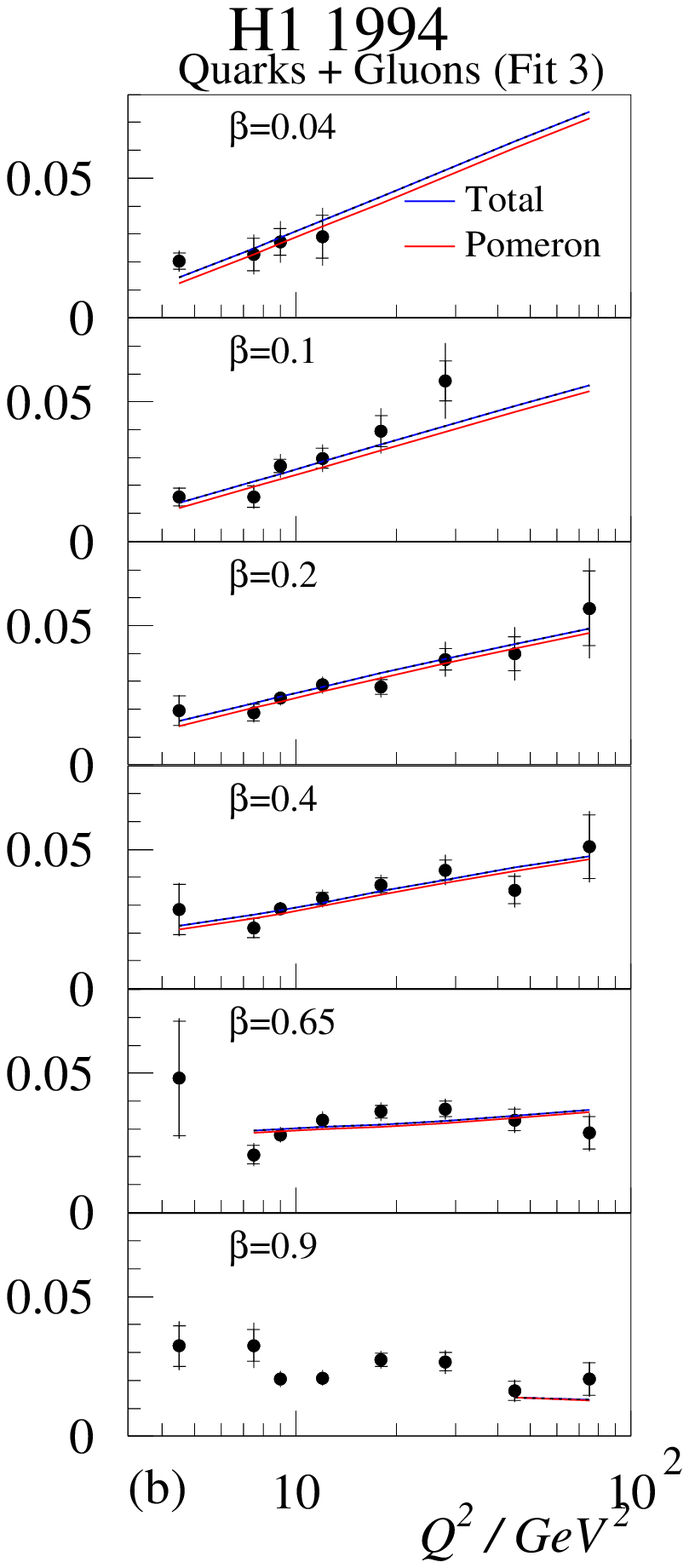,width=0.4\linewidth,clip=}}

\vspace*{-8.5cm}

\hspace*{6.cm}
\mbox{\epsfig{file=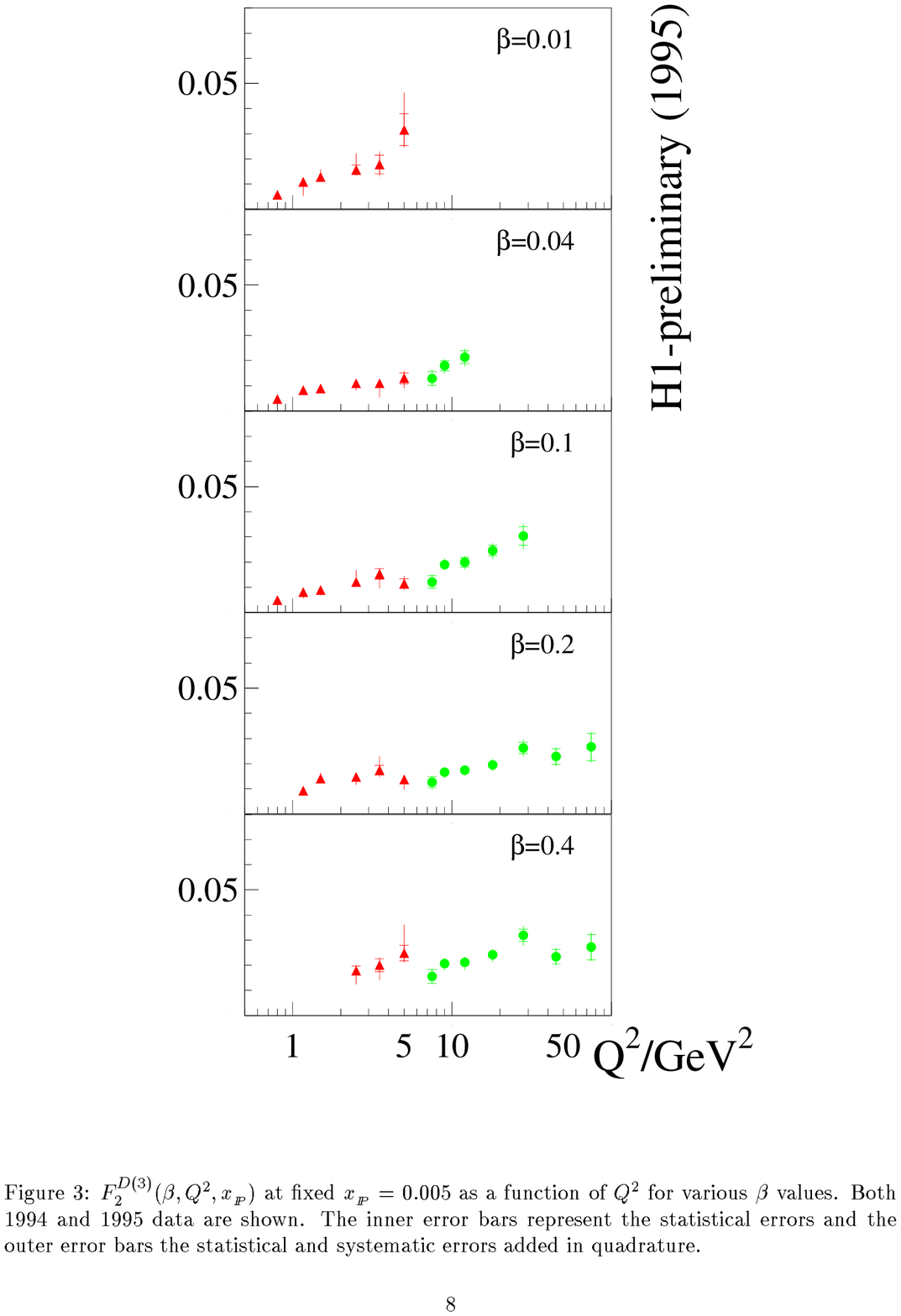,width=0.4\linewidth,clip=}}

\caption[~]{\label{scaling} \small Scaling violation of the
diffractive structure function $\xi F_2^D(\xi,\beta,Q^2)$ at
fixed $\xi=0.003$ (a) and $\xi=0.005$ (b) vs $Q^2$. The solid
line is a DGLAP fit.}        
\end{figure}

The diffractive structure function shows scaling violation as
depicted in Figure \ref{scaling}.
The structure function rises with $Q^2$ up to very high $\beta=0.65$
values \cite{H1-Diff,H1-1995}. The H1-Ansatz for $F_2^{D(3)}$ 
has been taken and the two
structure functions $F_2^{\Pom}$ and $F_2^M$ parametrised at a fixed
$Q_0^2=3GeV^2$ through parton distributions. These parton distributions
are evolved according to Dokshitzer-Gribov-Lipatov-Altarelli-Parisi (DGLAP)
\cite{DGLAP} equation to any other $Q^2$ and compared
with the data. For the meson piece, $F_2^M$, a pion structure function
was taken. For the Pomeron piece the ratio of quarks and gluons at the
starting scale $Q^2_0$ and its corresponding shapes were fitted with the
result that 90\% of the Pomeron momentum is carried by gluons at
$Q^2=4.5~GeV^2$ and still 80\% at $Q^2=75~GeV^2$. 

\subsection{Hard Diffraction at the Tevatron}
Hard diffraction has also been observed in $p\bar{p}$ collisions at the
Tevatron where the hardness scale is given by the required jets in the
final state or the production of a heavy gauge boson like $W$. Three
processes are considered:
 
1. $ p+\bar{p} \rightarrow p+gap+(X \rightarrow jet1+jet2+X')$.
The kinematics has been defined in Figure \ref{diff-picture}a.
Without the measurement of the final state proton the kinematical
variable $\xi$ cannot be measured. Its distribution is taken from MC
model calculations (POMPYT) with a factorised Pomeron flux
$f_{\Pom/p}(\xi,t)=\frac{K}{\xi^{2\alpha_{\Pom}(t)-1}}F^2(t)$
and a hard Pomeron structure function $\beta G(\beta)=6\beta(1-\beta)$ \cite{Schlein}.
In this model $\xi$ is restricted  through the final state topology to
$0.005 \le \xi \le 0.015$. The ratio of diffractive dijets to all
dijets with $E_T>20GeV$ is  $R_{JJ}=0.75 \pm 0.05(stat)\pm0.09(syst) \%$
(CDF \cite{CDF-jet}), and $0.67\pm0.05(stat)\%$ (D0). This ratio is sensitive  to the
relative ratio of quarks and gluons in the Pomeron.

2. $p+\bar{p} \rightarrow p+(W \rightarrow e \nu +X')$.
The gluon content in the Pomeron does not contribute much (suppressed by
$\alpha_s$). The result from CDF \cite{CDF-W} for the ratio of diffractive to
non-diffractive W-production $R_W=1.15\pm0.55(stat)\pm
0.20(syst) \%$.

\begin{figure}[htbp] 

{\bf \Large a)} \hspace*{10cm} {\bf \Large b)}

\begin{center}

\vspace*{-1.0cm}
\mbox{\epsfig{file=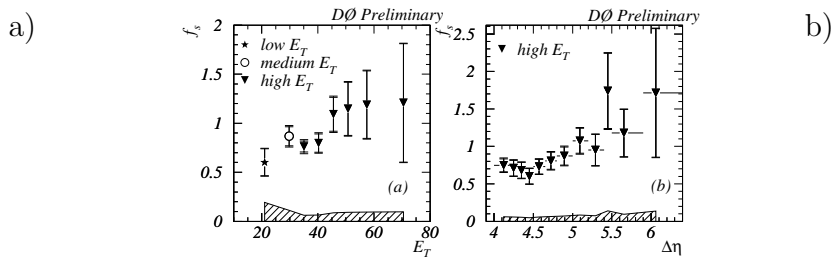,width=0.62\linewidth,clip=}}
\end{center}
\vspace*{-0.5cm}
\caption[~]{\label{D0} \small Fraction $f_s$[\%] of events with
a rapidity gap between jets vs jet $E_T$ (a) and vs the gap size
$\Delta \eta$ (b).}         
\end{figure}

3. $p+\bar{p} \rightarrow jet1+gap+jet2$.
Rapidity gaps of size $\Delta \eta$ between jets from fluctuations in
fragmentation are exponentially suppressed. Figure \ref{D0} shows data from D0 \cite{D0}
where the
fraction of events  with a large gap as a function of the jet $E_T$
(fig. \ref{D0}a) and as a function of $\Delta \eta$ (fig. \ref{D0}b)
are plotted.

\vspace*{-0.3cm}

It has become clear during the year, that not all 3 types of observations
can be explained by this simple factorised POMPYT model. It remains to be
seen if factorisation could be restored as in the case of HERA.

\section{Fragmentation}
\subsection{Charged Multiplicity in Diffraction at HERA}
Fragmentation gives the link between cross section on the parton level
and observed particles in the detector.
The charged particle multiplicity grows with the CM energy, with
$\sqrt{s}$ in $e^+e^-$ annihilation and with $W$ in lepton-proton
scattering. It has been shown that the multiplicity is independent of
$Q^2$ \cite{H1-mult}. The question arises if the diffractive system 
$X$ defined in
Figure \ref{diff-picture}b fragment as an independent object, i.e. is the
multiplicity in diffraction governed by $M_X$ or still by $W$? 

\begin{figure} 
\vfill \begin{minipage}{.46\linewidth}
\begin{center}
\mbox{\epsfig{file=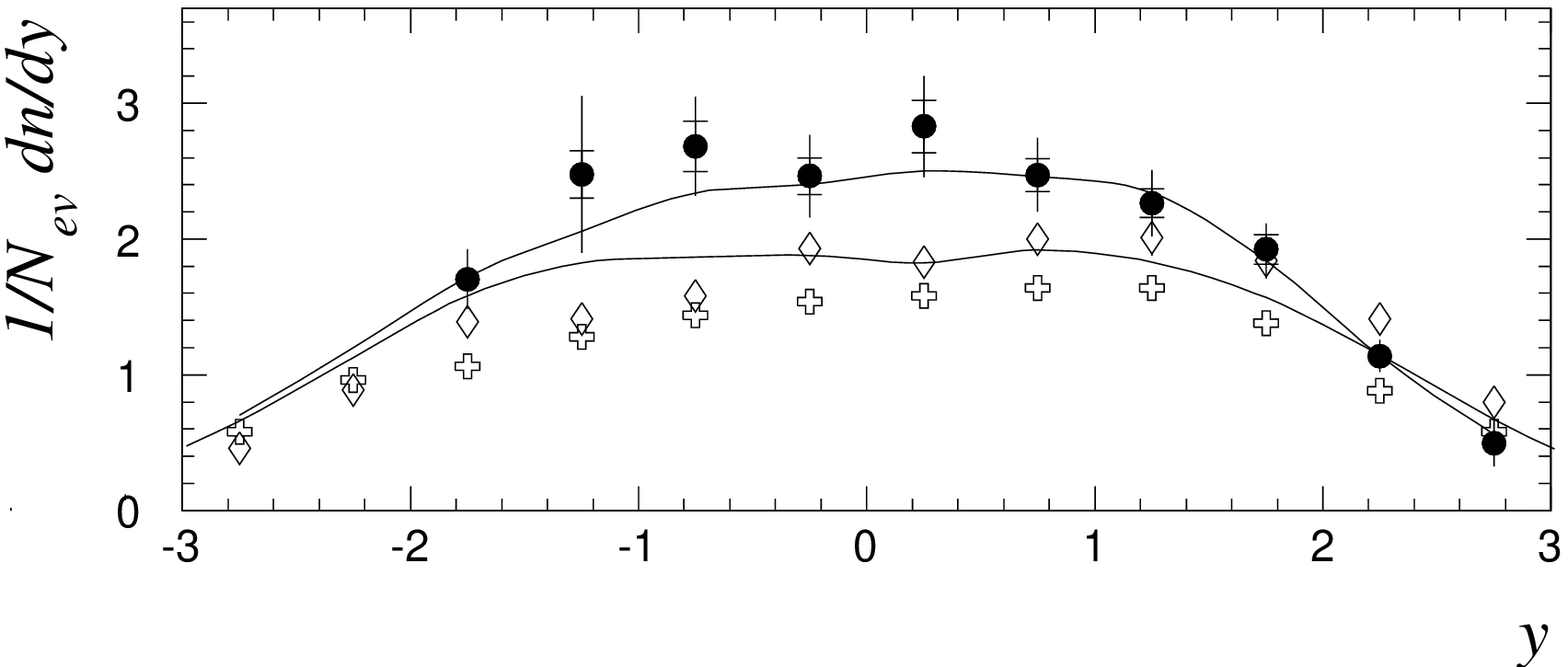,width=\linewidth,clip=}}
\end{center}
\vspace*{-0.5cm}

\caption[~]{\label{multi} \small Particle density $1/N dn/dy$ 
as a function of rapidity for $M_X=22.5~GeV$ (full symbols, H1), for $W=19~GeV$
(open crosses, EMC $\mu p$) and for $W=23.4~GeV$ (open circle, E665 $\mu D$).
The full curve is the RAPGAP simulation for $\gamma p$ processes and the dashed line from JETSET ($e^+e^-$).}
\end{minipage}\hfill
\begin{minipage}{.51\linewidth}
\begin{center}
\mbox{\epsfig{file=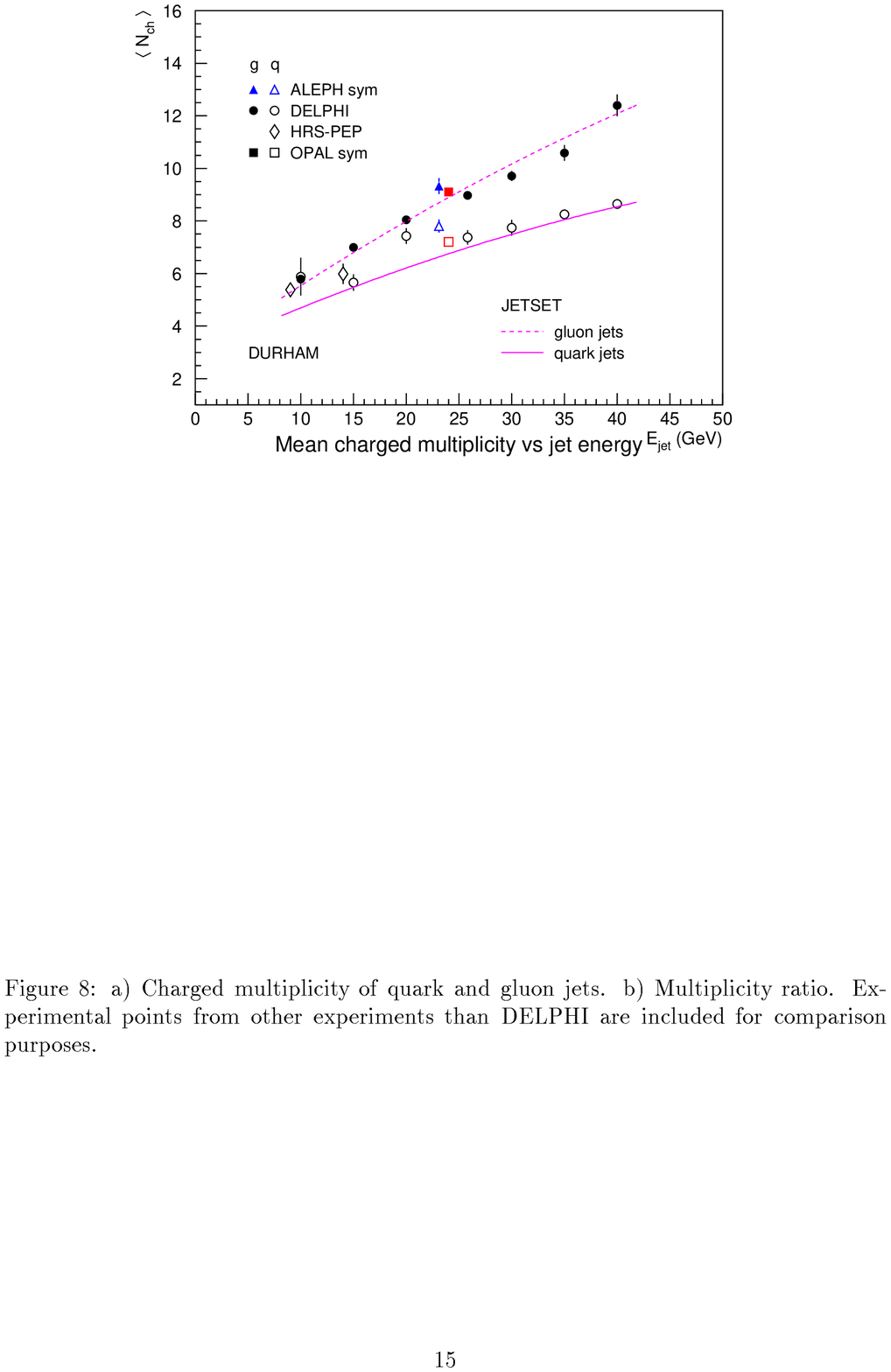,width=1.04\linewidth,clip=}}
\end{center}

\vspace*{-0.5cm}

\caption[~]{\label{quark-gluon} \small Mean charged
multiplicity vs jet energy for gluon jets (full symbols) and
quark jets (open symbols).}
\end{minipage}          
\end{figure}

In Figure \ref{multi} we
compare the particle density $\frac{1}{N}\frac{dn}{dy}$ at the plateau 
\cite{H1-diff-mult} as
a function of $M_X$ and compare it to inclusive (non-diff) DIS at $W=M_X$.
The multiplicity in diffractive events is about 0.6 units higher. We conclude
that hadronisation in diffraction cannot be described  by the
fragmentation of a single colour string as implemented in the Lund model.
The dominant diagram  according to \cite{Buchmueller} is the one in Figure
\ref{Feynman}c where the system $X$ is rather a colour octet string.

\subsection{Charged Multiplicities in Quark and Gluon Jets}
One expects a larger multiplicity in gluon jets than in quark jets
because of the stronger effective $g\rightarrow gg$ than $q \rightarrow
qg$ coupling. Gluon initiated jets are selected at LEP in a 3-jet $q\bar{q}g$ 
topology, where the jets with identified heavy quarks are taken as quark 
initiated jets. The effect of a larger multiplicity of the remaining gluon 
jets (75-90\% purity) is clearly seen in Figure \ref{quark-gluon} where
results from all four LEP experiments are shown \cite{LEP}.

\subsection{Leading Particles in Fragmentation}
\begin{figure}[htbp]   
\vspace*{-0.7 cm}
\begin{center}
\mbox{\epsfig{file=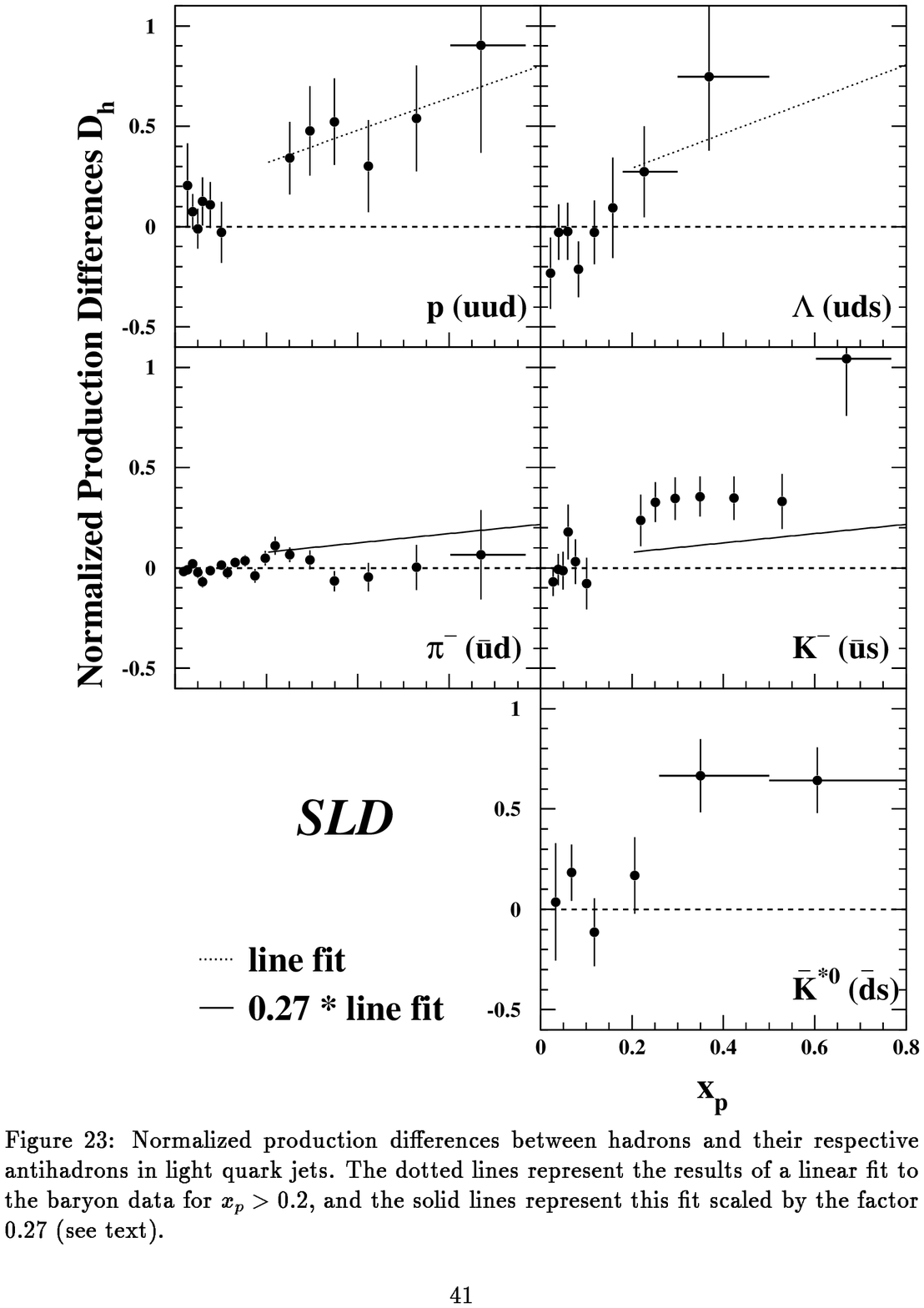,width=0.67\linewidth,clip=}}
\end{center}
\vspace*{-0.8cm}
\caption[~]{\label{leading} \small Normalized production differences $D_h$
between particles in quark jets and particles in antiquark jets vs $x_p$.
The dotted lines represent the results of a linear fit to the baryon data for $x_p>0.2$, and the solid lines represent this fit scaled by a dilution factor 0.27 valid for pions.}
\vspace*{-0.3cm}
\end{figure}

Leading particles carry quantum numbers such as flavour and spin of the
primary produced partons. This is demonstrated 
for flavours in a result of SLD (Figure
\ref{leading}) and for spin in LEP results.

The polarised initial state in $e^+e^- \rightarrow q \bar{q}$ at SLC
allows the distinction of a quark  initiated  jet as opposed  to
an antiquark jet. Consider jets from light flavours $u,d,s$ and let
$N(q\rightarrow h)$ be the number of hadrons of type $h$ from quark jets.
The production rate $R$ and the difference $D$ are defined
 $$R^q_h=\frac{1}{2N_{evts}}\frac{d}{dx_p}
\left[N(q \rightarrow h)+N(\bar{q}\rightarrow \bar{h})\right],
~~~~~D_h=\frac{R^q_h-R^q_{\bar{h}}}{ R^q_h+R^q_{\bar{h}}}.$$
Figure \ref{leading} plots $D_h$ vs $x_p$ \cite{SLD}. One sees 
 clear evidence for
production of leading baryons at high $x_p$. Also $K^-$ and
$\overline{K^*}$ are dominant over their antiparticles at high $x_p$.

Quarks  from $Z^0 \rightarrow q \bar{q}$ are polarised. It has been known
since several years, that this polarisation is transferred to the first
rank baryons $\Lambda,\Lambda_c,\Lambda_B$. LEP
measured a $\Lambda$-polarisation of $p_{\Lambda}=-32.9 \pm 7.6
\%$(OPAL \cite{Opal2}) and $=-32 \pm 7 \%$ (ALEPH). This is 
explained if the s-quark
carries the spin of the $\Lambda$ particle.

In $J^P=1^-$ vector mesons any spin alignment must arise from
hadronisation. The spin density matrix $\rho_{ij}$ can be measured
through angular dependence of the decay mesons with respect to the spin
quantisation axis  
$W(\cos\theta_H)=\frac{3}{4}\left[(1-\rho_{00})+(3\rho_{00}-1)\cos^2
\theta_H \right]$. 

In a statistical model \cite{Spin} 
$\rho_{00}=\frac{1}{2}(1-P/V)$ with $P/V$ ratio of 
pseudoscalar to vector mesons in fragmentation. In this model 
$\rho_{00}\le 0.5$ and $\rho_{00}=0$ for $P/V=1$, 
$\rho_{00}=1/3$ for $P/V=1/3$.

Results from CLEO, HRS, TPC-2$\gamma$, ALEPH, DELPHI, OPAL
give  $\rho_{00}\sim1/3$ for $\rho,D^*,B^*$-mesons,
 $\rho_{00}>1/3$ for $D^*,\Phi,K^*$-mesons at large fractional momentum,
and  even $\rho_{00}=0.66 \pm 0.11$ for $K^*$ at $x_p >0.7$ \cite{OPAL}.
A value of $\rho_{00}>0.5$ is an interesting observation and 
is not explained in any fragmentation model.

\vspace*{-0.4cm}

\section{Summary} 
Multi-dimensional distributions in diffraction
at the Tevatron and the precision data of HERA help the understanding 
of diffractive phenomena.
The semi-inclusive cross sections rise
with energy proportional  $(W^2)^{\lambda}$ where $\lambda$ grows with the
hard scale involved. 
Inelastic diffraction using only Pomeron exchange breaks
factorisation, both at HERA and the Tevatron. A sum of Pomeron 
exchange (80\% gluons) and meson exchange (quark-dominated exchange) 
restores factorisation at HERA.
The particle density $dn/dy$ at fixed $M_X$ is larger in diffractive DIS 
than in non-diffractive  DIS at the corresponding $W_{\gamma p}$
 which is related 
to a more complex colour string in diffraction than in non-diffractive DIS.
Fragmentation properties of gluon jets are well measured at LEP and 
show a larger charged multiplicity in gluon initiated jets than in quark jets.
Large spin alignment of leading vector mesons at Z$^0$ is observed, which
is an interesting observation and not explained so far.

\vspace*{-0.5cm}


\end{document}